\documentclass[a4paper,fleqn]{cas-dc}
\usepackage[numbers]{natbib}
\usepackage{subfigure}
\usepackage{enumerate}
\def\tsc#1{\csdef{#1}{\textsc{\lowercase{#1}}\xspace}}
\tsc{WGM}
\tsc{QE}
\tsc{EP}
\tsc{PMS}
\tsc{BEC}
\tsc{DE}

\begin{document}
\let\WriteBookmarks\relax
\def\floatpagepagefraction{1}
\def\textpagefraction{.001}
\shorttitle{R2MNet}
\shortauthors{xxx et~al.}

\title [mode = title]{Interpretative Computer-aided Lung Cancer Diagnosis: from Radiology Analysis to Malignancy Evaluation}                      



\author[1]{Shaohua Zheng}[style=chinese]
\fnmark[1]


\author[1]{Zhiqiang Shen}[style=chinese]

\fnmark[2]


\author[1]{Chenhao Pei}[style=chinese]
\fnmark[3]

\author[1]{Wangbin Ding}[style=chinese]
\fnmark[4]

\author[1]{Haojin Lin}[style=chinese]
\fnmark[5]

\author[2]{Jiepeng Zheng}[style=chinese]
\fnmark[6]

\author[1]{Lin Pan}[style=chinese]
\cormark[1]
\fnmark[7]

\author[2]{Bin Zheng}[style=chinese]
\cormark[2]
\fnmark[8]

\author[1]{Liqin Huang}[style=chinese]
\fnmark[9]

\address[1]{College of Physics and Information Engineering, Fuzhou University, Fuzhou 350108, China}

\address[2]{Thoracic Department, Fujian Medical University Union Hospital, Fuzhou 350001, China}

\cortext[cor1]{Corresponding author at: College of Physics and Information Engineering, Fuzhou University, Fuzhou 350108, China
E-mail: \href{mailto:panlin@fzu.edu.cn}{panlin@fzu.edu.cn}).}

\cortext[cor2]{Corresponding author at: Thoracic Department, Fujian Medical University Union Hospital, Fuzhou 350001, China
E-mail: \href{mailto:lacustrian@163.com}{lacustrian@163.com}).}

\begin{abstract}
Background and Objective: Computer-aided diagnosis (CAD) systems promote diagnosis effectiveness and alleviate pressure of radiologists. A CAD system for lung cancer diagnosis includes nodule candidate detection and nodule malignancy evaluation. Recently, deep learning-based pulmonary nodule detection has reached satisfactory performance ready for clinical application. However, deep learning-based nodule malignancy evaluation depends on heuristic inference from low-dose computed tomography (LDCT) volume to malignant probability, which lacks clinical cognition. 
\\
Methods: In this paper, we propose a joint radiology analysis and malignancy evaluation network (R2MNet) to evaluate the pulmonary nodule malignancy via radiology characteristics analysis. Radiological features are extracted as channel descriptor to highlight specific regions of the input volume that are critical for nodule malignancy evaluation. In addition, for model explanations, we propose channel-dependent activation mapping (CDAM) to visualize the features and shed light on the decision process of deep neural network (DNN).
\\
Results: Experimental results on the LIDC-IDRI dataset demonstrate that the proposed method achieved area under curve (AUC) of $96.27\%$ on nodule radiology analysis and AUC of $97.52\%$ on nodule malignancy evaluation. In addition, explanations of CDAM features proved that the shape and density of nodule regions were two critical factors that influence a nodule to be inferred as malignant, which conforms with the diagnosis cognition of experienced radiologists.
\\
Conclusion: Incorporating radiology analysis with nodule malignant evaluation, the network inference process conforms to the diagnostic procedure of radiologists and increases the confidence of evaluation results. Besides, model interpretation with CDAM features shed light on the regions which DNNs focus on when they estimate nodule malignancy probabilities.
\end{abstract}




\maketitle

\section{Introduction}
Lung cancer is the most common cause of cancer death worldwide \cite{cancer2015}. Lung cancer screening using low-dose computed tomography (LDCT) scans has been proved as an effective tool to reduce patient mortality \cite{lungCTsurvey}. However, A thorough inspection of a CT scan usually takes a radiologist around 10 minutes and diagnosis results are influenced by the doctor’s experience and emotion. With the increasing number of CT images, the data volumes to be analyzed overwhelm the capacity of radiologists. Computer-aided diagnosis (CAD) systems have the potential to reduce this burden. In recent years, deep learning-based methods have demonstrated impressive performance in medical image processing, and taken up a dominant position in the design of CAD systems \cite{unet, 3dunet, vnet, deeplung, dsb2017first, 3dprobabilisticCAD}.

\begin{figure}
\centering
\includegraphics[width=2.5 in]{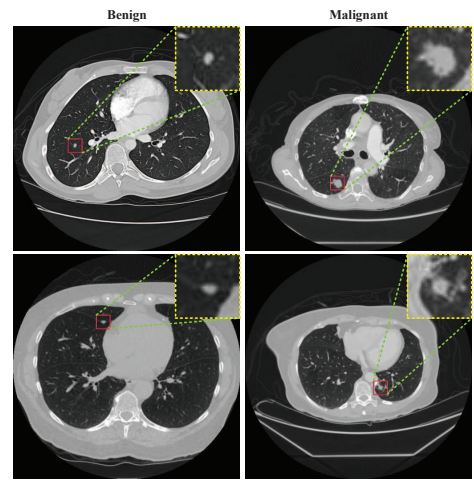}
\caption{Examples of benign (the left column) and malignant nodules (the right column). The red rectangles emphasize the nodule locations and the yellow dashed rectangles highlight the nodule areas. Figure best viewed in color.}
\label{fig1}
\end{figure}

\begin{figure}
\centering
\includegraphics[width=2.5 in]{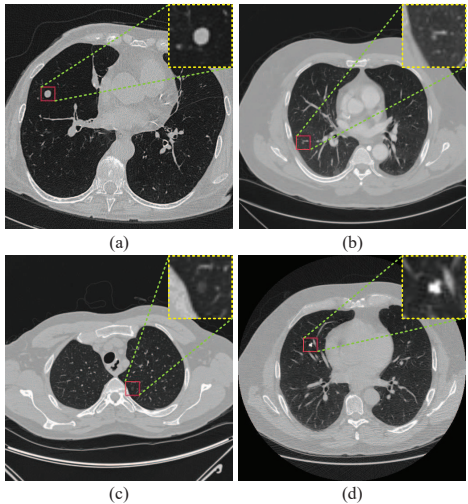}
\caption{Examples of nodules labeled as solid nodule (a), mix ground-glass opacity nodule (b), ground-glass opacity nodule (c), and calcified nodule (d). The red rectangles emphasize the nodule locations and the yellow dashed rectangles highlight the nodule areas. Figure best viewed in color.}
\label{fig2}
\end{figure}

A general deep learning-based CAD system for lung cancer diagnosis includes 1) a pulmonary nodule detection module that detects candidate pulmonary nodules, and 2) a nodule malignancy evaluation module that diagnoses the suspicious nodules proposed by the previous stage. Deep learning-based nodule detection has achieved remarkable results. However, deep learning-based nodule malignancy evaluation models that straightforwardly predict malignant probabilities are short of explanations of which regions deep neural networks (DNNs) focus on \cite{nodule_resnet, nodule_nonlocal}.
Doctors estimate nodule malignant risk mainly according to the shape and density of the nodules as well as other pathology information. Qualitatively, compared to the benign nodules, the malignant ones often have larger volumes, varied density, and irregular shapes. Examples of benign and malignant nodules are illustrated in Fig.\ref{fig1}. 
The inference results of the DNNs, therefore, lack confidence and interpretation.

To overcome the problems mentioned above, we propose a joint radiology analysis and malignancy evaluation network (R2MNet) that evaluates nodule malignancy according to radiology analysis. Specifically, radiology analysis aims to classify nodules as solid nodules (SN), ground-glass opacity nodules (GGO), mix GGO nodules (MGGO), and calcified nodules (CN) as shown in Fig.\ref{fig2}. The purpose of malignancy evaluation is to estimate malignant risk of nodule. R2MNet consists of two sub-networks, the radiology analysis network (RNet) and the malignancy evaluation network (MNet) to implemented these two task, respectively. To consolidate the two sub-networks, we design assisted gating units (AGUs) embedded in the MNet to transform the feature maps extracted by RNet as a channel descriptor to capture channel dependencies of that by MNet. Moreover, model interpretability is crucial in CAD. To enable our model explainable, we propose channel-dependent activation mapping (CDAM) that adopts channel dependencies of activation maps themselves for features interpretation. 
Extensive experiments on LIDC-IDRI \cite{lidc_idri} via 
five-fold cross-validation demonstrate that the proposed R2MNet achieves satisfactory performance on nodule malignancy evaluation. Moreover, its inference process conforms to clinical diagnosis procedure which increases the confidence level of evaluation results. Our contributions can be summarized as follows:
\begin{itemize}
    \item We propose R2MNet that integrates two sub-networks (RNet and MNet) to inference malignant risk via radiology analysis. The RNet extracts radiological feature using new labeled data. MNet evaluates nodule malignancy.
    
    \item To conjoin the two sub-networks of R2MNet, we design the AGUs embedded in MNet to transform the feature maps extracted by RNet as a channel descriptor to capture channel dependencies of that by MNet.
    
    \item To enable our model interpretable, we propose CDAM that exploits channel dependencies of the activation maps for visualizing explanation.
    
    \item  Extensive experiments on the LIDC-IDRI dataset indicate that our method achieves promising accuracy for nodule malignancy evaluation. Remarkably, the inference process conforms to clinical diagnosis procedure.
\end{itemize}

The rest of this paper is organized as follows. In Section \ref{sec:relatedwork}, we review the relevant literature. Datasets and their corresponding preprocessing are specified in Section \ref{sec:materials}. Section \ref{sec:methods} elaborates on the proposed methods. Experiments setting and results are shown in Section \ref{sec:experiments_and_results}. In Section \ref{sec:discussion}, we discuss the experiment results and analyze the superiority and limitations of our approach. Section \ref{sec:conclusion} concludes this paper.

\section{Related work}
\label{sec:relatedwork}
In the following, we review the works related to pulmonary nodule classification, long-range dependencies, and Class Activation Map (CAM)-based explanation.

\subsection{Pulmonary nodule classification}
\label{subsec:nodule_classification}
In a deep learning-based CAD system, nodule classifiers either reduce false-positive nodules following nodule detectors or evaluate nodule malignancy in the back of the CAD systems. Setio et al. extracted 2D patches from nine symmetrical planes of a cube for false positive reduction \cite{multiviewCNN}. Dou et al. encoded multi-level context information with 3D Convolutional Neural Network (CNN) to reduce false positives \cite{MuitiLevel3DCNN}. MD-NDNet integrated nodule volumetric information and spatial nodule correlation features from sagittal, coronal, and axial planes to decrease false positive rate \cite{MDNDNet}. Winkels et al. developed a 3D version of group equivariant convolutional networks that generalizes automatically over discrete rotations and reflection for false-positive reduction \cite{NoduleEquivariantCNN}. False-positive reduction using CNNs that identifies input CT volumes whether have nodules or not conforms to the clinical basis. However, nodule benign/malignant evaluation directly from CT to malignant probability lacks interpretation of features extracted by CNN \cite{nodule_resnet, nodule_nonlocal}. To improve model interpretability, Hussein et al. adopted multiple CNNs based on graph regularized sparse multi-task learning for malignant risk stratification \cite{hussein2017risk}. Similarly, Wu et al. integrated the tasks including classification and segmentation in a multi-task learning manner \cite{wu2018joint}. 
In this work, we exploit radiological features as a channel descriptor for nodule malignancy evaluation. Besides, we employ the proposed CDAM for model explanation. Overview of the proposed model is introduced in Section \ref{subsec:R2MNet}. 

\subsection{Long-range dependencies}
\label{subsec:Long_range_dependencies}
Learning long-range dependencies is of great importance in deep neural networks. Long-range dependencies enable networks to capture large receptive field and learn global features. Convolutions are local operations in which long-range dependencies can only be captured when these operations are applied repeatedly. The transformer was one of the first attempts to apply a self-attention mechanism to model long-range dependencies in machine translation \cite{transformer}. Non-local operation captured the pixel-level pairwise relations for solving computer vision  \cite{nonlocal}. GCNet improved the Non-local network with less computation while maintained the effectiveness of long-range dependencies capturing \cite{gcnet}. To learn channel-wise dependencies of feature maps, SENet recalibrated the channel dependency with global context features as each channel of feature maps corresponding to the specific region of the input image \cite{senet}. Motivating by the superiority of SENet, we propose a AGU for recalibrating channel relationships using specific features as a channel descriptor. Details of the AGU are presented in Section \ref{subsec:AGU}.

\subsection{CAM-based Explanation}
\label{CAM_explanation}
Activation maps visualization has been the most mainstream method for CNN interpretation. Specifically, the Class Activation Map (CAM) is one of the widely adopted methods \cite{CAM}. CAM-based explanations provide feature visualization for explanations with a weighted combination of activation maps learned from CNN \cite{CAM, GradCAM, GradCAM++, ScoreCAM}. 

CAM identified discriminative regions by a linear weighted combination of activation maps of the last convolutional layer before the global pooling layer \cite{CAM}. However, it is only appropriate for a restricted class of CNNs that contain global average pooling layers and fully connected layers. To extend the range of application of CAM, Grad-CAM generalized the definition of the weighting coefficients as the gradient of class confidence concerning the activation map and applies to a significantly broader range of CNN model families \cite{GradCAM}. The variation of Grad-CAM, Grad-CAM++ aimed to provide better localization of objects as well as explaining occurrences of multiple objects of a class in a single image \cite{GradCAM++}. Using gradient to incorporate the importance of each channel towards the class confidence is a natural choice. The gradient information for a deep neural network can be noisy and also tends to vanish due to saturation in sigmoid or the flat zero-gradient region in Rectified Linear Unit (ReLU). Instead of using the gradient information flowing into the last convolutional layer to represent the importance of each activation map, Score-CAM exploited the importance as the linear combination of score-based weights and activation maps \cite{ScoreCAM}. However, the aforementioned methods adopted weighting coefficients derived from external data, which may introduce noise and bias. Therefore, we propose the CDAM for activation maps visualization where the weighting coefficients are calculated from the activation maps themselves. Details of the CDAM are presented in Section \ref{subsec:CDAM}.

\section{Materials}
\label{sec:materials}
In this section, we introduce the database used in our experiments. Data annotation and preprocessing methods are also specified.
\subsection{Dataset}
\label{subsec:dataset}
In this study, we use a selected version of the LIDC database \cite{lidc_idri} provided in the LUNA16 challenge \cite{luna16} which consists of $888$ CT scans comprising a total of $1186$ nodules. We have obtained the nodules malignancy from the annotation files in the LIDC-IDRI dataset. Nodules with an average score higher than $3$ were labeled as malignant and lower than $3$ are labeled as benign. Some nodules were removed from the experiments in the case of the averaged malignancy score $3$, ambiguous IDs, and rated by only one or two radiologists, which resulted in a total of $1004$ nodules where there were $450$ malignant nodules and $554$ benign nodules. 

\subsection{Radiological categories annotation}
\label{subsec:radiological_categories_annotation}
For nodule radiological analysis, two experienced radiologists labeled nodules as SN, GGO, MGGO, CN according to the 3D radiological features of LDCT scans using ITK-SNAP \cite{ITKSnap}. The steps of data annotation are briefly listed as follows.
\begin{enumerate}[1)]
    \item Two experienced doctors, respectively, marked the class of all nodules based on radiological characteristics.
    \item Then, they carefully inspected and corrected the labeled results, respectively.
    \item The final version was obtained by discussing and remark the different classes labeled in previous steps.
\end{enumerate}

\subsection{Preprocessing}
The data preprocessing follows four steps.
\begin{enumerate}[1)]
    \item  \textbf{Normalization}. We clipped hounsfield units (HU) of the raw CT data into [-1200, 600] and normalized them into [0, 1].
    \item \textbf{Extraction}. Foreground regions of normalized CT scans were extracted according to the ground truth masks provided by LUNA 16 challenge.
    \item \textbf{Resample}. We resampled all CT volumes to have 1 mm spacing in the z-, y-, x-dimension.
    \item \textbf{Crop}. The nodule regions used to train and test our method were cropped according to the experiment configurations (i.e. 2D/3D format and size)
\end{enumerate}

\begin{figure*}
\centering
\includegraphics[width=6.5 in]{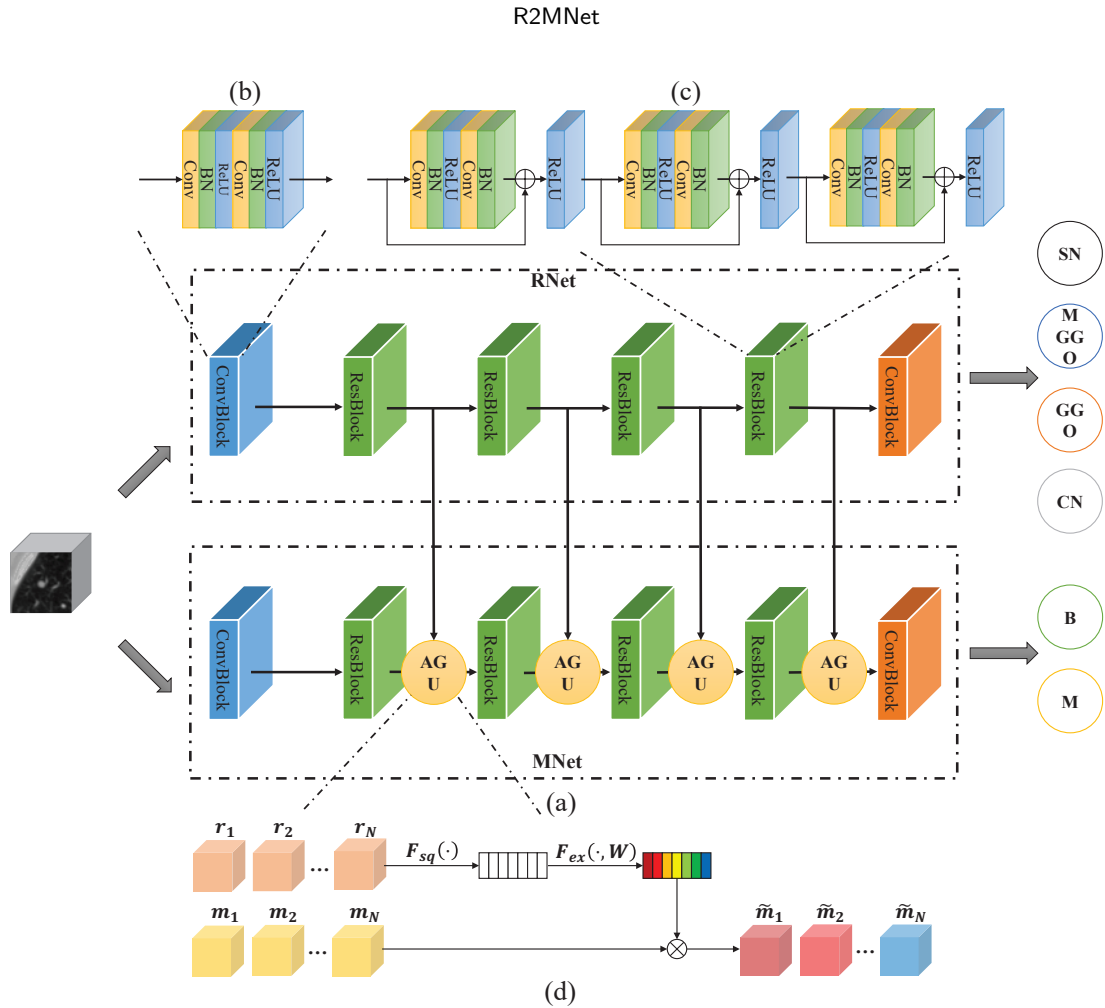}
\caption{The diagram of the proposed method. (a) R2MNet. Note that we omit the four max pooling layers each of which is behind the Residual blocks for illustration convinience. (b) The convolutional block. (c) The residual blocks. (d) The AGU module.}
\label{fig3}
\end{figure*}

\section{Methods}
\label{sec:methods}
\begin{figure*}
\centering
\includegraphics[width=5.5 in]{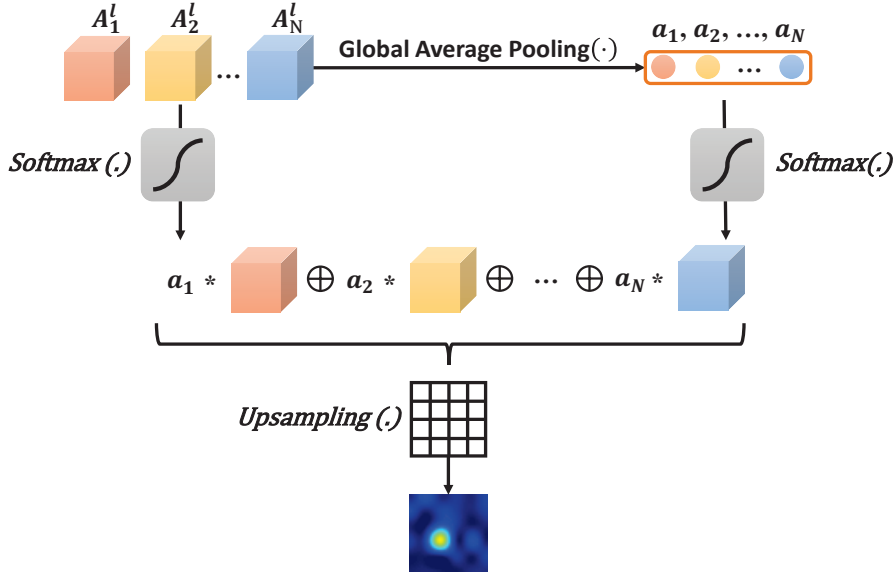}
\caption{The diagram of the proposed CDAM. Activation maps are linearly weighted to generate visual explanation.}
\label{fig_cdam}
\end{figure*}

In this section, we introduce our R2MNet and detail its components. Then, the implemented details are presented. The proposed R2MNet are shown in Fig.\ref{fig3}. The diagram of R2MNet is illustrated in Fig.\ref{fig3}(a). R2MNet is composed of two CNN trained in multi-task learning manner (Section \ref{subsec:R2MNet}). The AGU transforms radiological features into a channel descriptor to facilitate malignancy evaluation (Fig.\ref{fig3}(d)). CDAM are proposed for model explanation (Fig.\ref{fig_cdam}).

\subsection{R2MNet}
\label{subsec:R2MNet}
Here we present our R2MNet and provide an overview of the key components. The proposed R2MNet takes a 3D CT volume of as input and provides as outputs a radiology class and a nodule malignant score. Specifically, the R2MNet consists of two improved residual networks \cite{resnet}, i.e., RNet and MNet as illustrated in Fig.\ref{fig3} (a). MNet includes two convolutional blocks(Fig.\ref{fig3} (b)), four residual blocks each of which contains three residual units (Fig.\ref{fig3} (c)), four AGUs (Fig.\ref{fig3} (d)), and four max-pooling layers. The architecture of RNet is similar to MNet but without AGUs. The proposed method can combine nodule radiological features for nodule malignancy evaluation.
The RNet and MNet are trained simultaneously in a multi-task learning manner. This is different than current approaches that use directly a CNN for malignancy estimation \cite{nodule_resnet, nodule_nonlocal}. The goals of RNet are extracting radiological features of pulmonary nodules for nodule evaluation as well as providing the radiological characteristics as a reference for practice diagnosis. The outputs of RNet are four categories probabilities and radiological features. The radiological features are transformed into a channel descriptor by AGU (Section \ref{subsec:AGU}) to render the MNet focus on nodule area. MNet takes as inputs the CT volume data and the radiological features for pulmonary nodule malignancy evaluation. The loss function for training our networks is weighted cross-entropy (CE) loss.
\begin{equation}L(Y_r, Y_m, \hat{Y}_r, \hat{Y}_m)= \lambda L_{CE}(Y_r,\hat{Y}_r) + (1-\lambda) L_{CE}(Y_m,\hat{Y}_m)\end{equation},
where $Y_r$ and $Y_m$ are the ground truth, and $\hat{Y}_r$ and $\hat{Y}_m$ are predictions of the R2MNet.

\subsection{Assisted Gating Unit}
\label{subsec:AGU}
The vanilla SE layer \cite{senet} adopted the input features to capture channel dependencies in 2D senario. The AGU, instead, transforms the features extracted by RNet as a channel descriptor to capture channel dependencies of that by MNet in 3D senario. The diagram of AGU is shown in Fig. \ref{fig3} (d). Specifically, we transform radiological features into a channel descriptor to capture the channel dependencies of malignancy features. Similar to the SE block, we model channel interdependencies to recalibrate filter responses in two steps (i.e., squeeze and excitation) as discussed follows. 
\begin{enumerate}[1)]
    \item \textbf{Squeeze}. 
    Squeeze operations are adopted for global information embedding. In R2MNet, radiological features $R = [r_{1}, r_{2}, …, r_{N}]$ are squeezed to a channel descriptor by using global average pooling (GAP). Noting that more sophisticated aggregation strategies could be employed here as well, we adopt GAP as used in \cite{senet}. The channel descriptor $T = [t_{1}, t_{2}, …, t_{N}] \in \mathbb{R}^{C}$ is computed as:
    \begin{equation}
    \label{eq1}
    t_{n}=F_{sq}\left(r_{n}\right)=\frac{1}{D \times H \times W} \sum_{i=1}^{D} \sum_{j=1}^{H} \sum_{k=1}^{W} r_{n}(i, j, k)
    \end{equation}
    where $F_{sq}$ is the squeeze operation, and $D$, $H$ and $W$ denotes the depth, height and width of the feature maps.
    
    \item \textbf{Excitation}. 
    The following operation takes as input the information aggregated in the last step to capture channel dependencies, i.e. $S = [s_{1}, s_{2}, …, s_{N}] \in \mathbb{R}^{C}$. The excitation operation can be formulated as follows:
    \begin{equation}
    S=F_{ex}(T, W)=\sigma(g(T, W))=\sigma\left(W_{2} \delta\left(W_{1} T\right)\right)
    \end{equation}
    where $\sigma$ is the sigmoid function and $\delta$ refers to ReLU activation, $W_{1} \in \mathbb{R}^{\frac{C}{r} \times C}$ and $W_{1} \in \mathbb{R}^{C \times \frac{C}{r}}$. Similar to \cite{senet}, we form a bottleneck including a dimensionality-reduction layer with parameters $W_1$ with reduction ratio $r$, a ReLU acitivation, and then a dimensionality-increasing layer with parameters $W_2$. Finally, the recalibrated malignant features $\widetilde{M} = [\widetilde{m}_{1}, \widetilde{m}_{2}, …, \widetilde{m}_{N}] \in \mathbb{R}^{C}$ are obtained by rescaling the malignant features $M = [m_{1}, m_{2}, …, m_{N}] \in \mathbb{R}^{C}$ with the radiological channel descriptors $T$:
    \begin{equation}
    \widetilde{m}_{n}=F_{\text {scale}}\left(t_{n}, s_{n}\right)=s_{n} \cdot t_{n} 
    \end{equation}
    where $F_{\text {scale}}$ denotes channel-wise multiplication.
\end{enumerate}

\subsection{Channel-Dependent Activation Mapping}
\label{subsec:CDAM}
We propose CDAM for 3D features visualization motivating by CAM-based methods, as shown in Fig.\ref{fig_cdam}. CAM is a technique for identifying discriminative regions by a linearly weighted combination of activation maps of the last convolutional layer before the global pooling layer \cite{CAM}. The motivation behind CAM is that each activation map of a CNN layer contains different spatial information about the input $X$ and the importance of each channel is the weight of the linear combination of the fully connected layer following the global pooling. However, if there is no global pooling layer or there is no fully connected layers, CAM will not apply due to no definition of the weighted coefficients. Grad-CAM \cite{GradCAM} and its variations \cite{GradCAM++} generalize CAM to models without global pooling layers by employing gradients as weights.

Instead of using weights of the fully connected layer or gradient information derived from external layers, CDAM employs activation maps themselves to obtain weights for a linear combination of activation maps. Formally, CDAM is defined as:
\begin{equation}
L_{CDAM} =\operatorname{ReLU}\left(\sum_{i=1}^{C} \alpha_{i} A_{i}^{l}\right)
\end{equation}
where $A^{l}$ denotes the activations of the $l$th CNN layer, $A_{i}^{l}$ refers to the activation map for the $i$th channel of $A^{l}$, and $a = [a_{1}, a_{2}, ..., a_{n}] \in \mathbb{R}^{C}$ is defined as:
    \begin{equation}
    a_{n} =\frac{1}{D \times H \times W} \sum_{i=1}^{D} \sum_{j=1}^{H} \sum_{k=1}^{W} A_{n}^{l}(i, j, k)
    \end{equation}, 
 We apply a ReLU activation to the linear combination of maps because we are only interested in the features that have a positive influence. Both $a_l$ and $A_l$ are utilized after the Softmax activation because the relative output value after normalization is more reasonable to measure the relevance than the absolute output value. Furthermore, to capture voxel-wise importance, we up-sample $L_{CDAM}$ to the input resolution using bicubic interpolation.

\subsection{Implemented details}
\label{subsec:implemented_details}
 The network were performed on PyTorch \cite{pytorch}. The models were trained via Adam optimizer \cite{adam} with standard back-propagation. Data augmentation operations i.e., scaling, flip, and rotation were also employed in the experiments. The learning rate was set as a fixed value of $1e-4$ and the number of epochs was $100$. The networks were trained on a single NVIDIA GeForce GTX 1080Ti.

\begin{figure*}
\centering
\includegraphics[width=5.5 in]{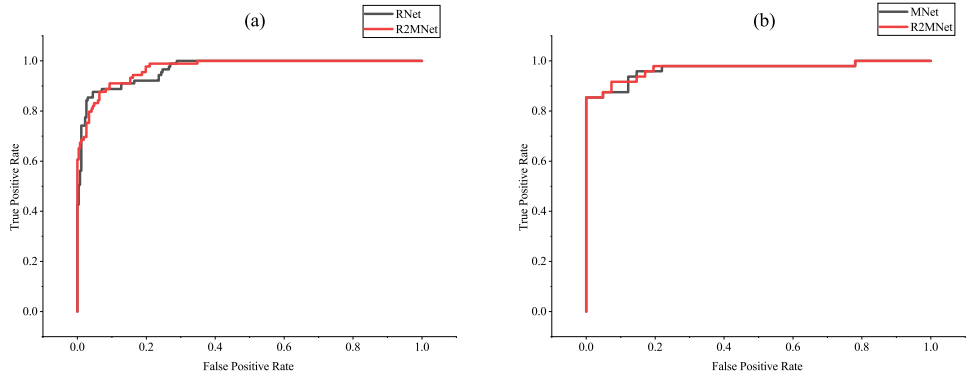}
\caption{
ROC curves of RNet and R2MNet on the radiology analysis (a), and malignancy evaluation (b).}
\label{fig4}
\end{figure*}

\begin{figure*}
\centering
\includegraphics[width=6.5 in]{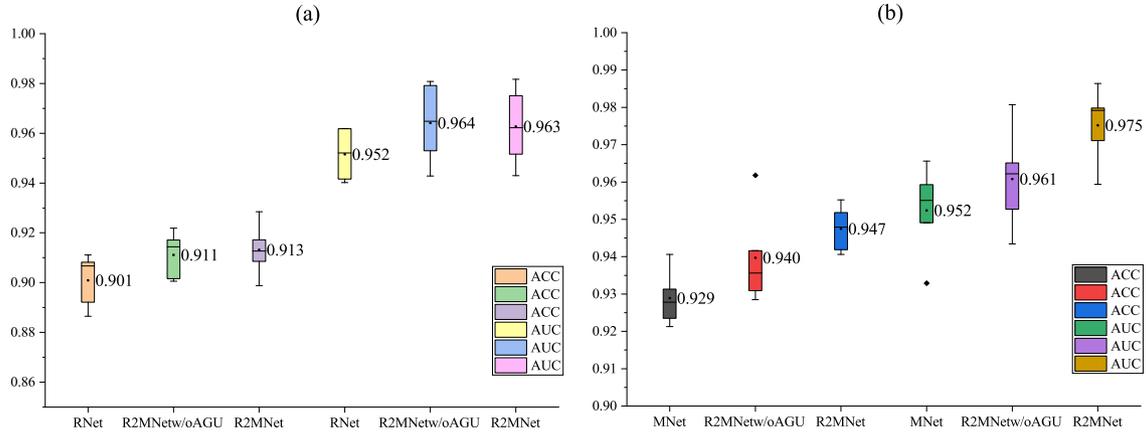}
\caption{Comparison among RNet, MNet, R2MNetw/oAGU, and R2MNet with accuracy and AUC on radiology analysis (a) and malignancy evaluation (b), respectively. The first three columns are the accuracy boxes and the remaining are AUC ones. Each scalar in the left of the corresponding boxes is the average value.}
\label{fig5}
\end{figure*}


\begin{table}[!t]
\centering
\caption{Performance comparison of RNet and R2MNet on radiology analysis.}
\begin{tabular}{llllll}
\hline
Model  & SN    & MGGO  & GGO   & CN    & AUC   \\ \hline
RNet   & 95.50 & 89.88 & 91.01 & 96.63 & 95.21 \\
R2MNet & 96.63 & 92.13 & 91.01 & 97.75 & 97.08 \\ \hline
\end{tabular}
\label{table2}
\end{table}

\begin{table*}[!t]
\centering
\caption{Performance comparison measured by accuracy and AUC ($mean\pm s.d.\%$) for MNet, R2MNetw/oAGU, and R2MNet on radiology analysis and malignancy evaluation.}
\begin{tabular}{lllll}
\hline
Task           & \multicolumn{2}{l}{Radiological analysis} & \multicolumn{2}{l}{Malignant   evaluation} \\ \hline
Model          & Accuracy             & AUC                  & Accuracy              & AUC                  \\ 
MNet & $90.82\%\pm1.09$         & $95.15\%\pm1.05$  & $92.89\%\pm0.76$        & $95.24\%\pm1.25$       \\ 
R2MNet\_w/oAGU  & $91.11\%\pm0.95$         & $96.41\%\pm1.64$  & $93.97\%\pm1.33$        & $96.08\%\pm1.40$    \\ 
R2MNet  & $91.13\%\pm1.10$         & $96.27\%\pm1.60$    & $94.74\%\pm0.62$        & $97.52\%\pm1.04$       \\ \hline 
\end{tabular}
\label{table3}
\end{table*}

\section{Experiments and results}
\label{sec:experiments_and_results}
In this section, we evaluate the proposed R2MNet on the LIDC-IDRI database and show the results. First, we performed nodule characteristics identification and malignancy evaluation individually. Then, we combined two tasks in multi-task learning where radiology analysis assisted malignancy evaluation. For model explanations, we visualized the feature maps and analyzed their characteristics. Experimental results show that the proposed method achieved higher performance compared to the baseline.


\subsection{Nodule Radiology analysis}
\label{subsec:radio_analysis}
Nodule radiology analysis aims to classify nodules as SN, MGGO, GGO, and CN nodules. Identifying these characteristics renders the model to learn radiological features for facilitating malignant evaluation. In addition, these characteristics can assist radiologists in determining nodule attributes as well. Experimental results of nodule characteristics classification are listed in Table \ref{table2}. Both RNet and R2MNet achieved accuracy higher than $90\%$ among the four categories. After combined with MNet, the performance of R2MNet either remained the accuracy level of RNet (GGO, CN) or was higher than that of RNet (SN, MGGO). Also, the area under curve (AUC) of R2MNet is larger than that of RNet. According to the Fig.\ref{fig4} (b), the ROC curve of R2MNet nearly surrounds that of the RNet.

\subsection{Nodule malignancy evaluation}
\label{nodule_malignancy_evaluation}
 Radiological features of pulmonary nodules can assist CNN for malignant classification because the inference procedure conforms to the diagnosis process. To testify the effectiveness of the proposed method, we conducted experiments of nodule malignant classification. As shown in Table \ref{table3}, R2MNet outperforms MNet with an accuracy gain of $1.72\%$ and an AUC gain of $1.38\%$, respectively. Moreover, the accuracy and AUC of R2MNet are more stable compared to MNet according to the standard deviation. The ROC curves of MNet and R2MNet are depicted in Fig.\ref{fig4} (c). To compare the overall performance of MNet and R2MNet through five-fold cross-validation, we also illustrated the box plots with accuracy and AUC in Fig.\ref{fig5}. As shown, compared to MNet, R2MNet achieved more stable and higher results.
 
\subsection{Ablation study}
\label{subsec:ablation_study}
We conducted an ablation study to investigate the individual contributions of R2MNet and AGU module. We implemented the experiments both on radiology analysis and malignant evaluation. The experiments were performed from two ends; on the one hand, we just included the radiology analysis in nodule malignant evaluation, which resulted in a fundamental version of R2MNet (i.e., R2MNetw/oAGU). On the other hand, the AGU modules were introduced into the preliminary R2MNet to construct the final version of the proposed method (i.e., R2MNet).

In nodule radiology analysis, a comparison was made among RNet, R2MNetw/oAGU, and R2MNet. As indicated in Table \ref{table3} the accuracy and AUC scores of the R2MNetw/oAGU are similar to that of R2MNet. Both of them slightly outperforms RNet. Results are shown in Fig.\ref{fig5}(a).

In nodule malignancy evaluation, a comparison was implemented among MNet, R2MNetw/oAGU, and R2MNet. The results of the 
five-fold cross-validation are listed in Table \ref{table3}. We can observe from the table that combining radiological analysis with malignant evaluation improves performance over doing the latter only. Further, when AGU is introduced into R2MNet, the synergy between these two components generates the best performance. The illustration of these results is shown in Fig.\ref{fig5} (b). 

\begin{figure*}
\centering
\includegraphics[width=5.5 in]{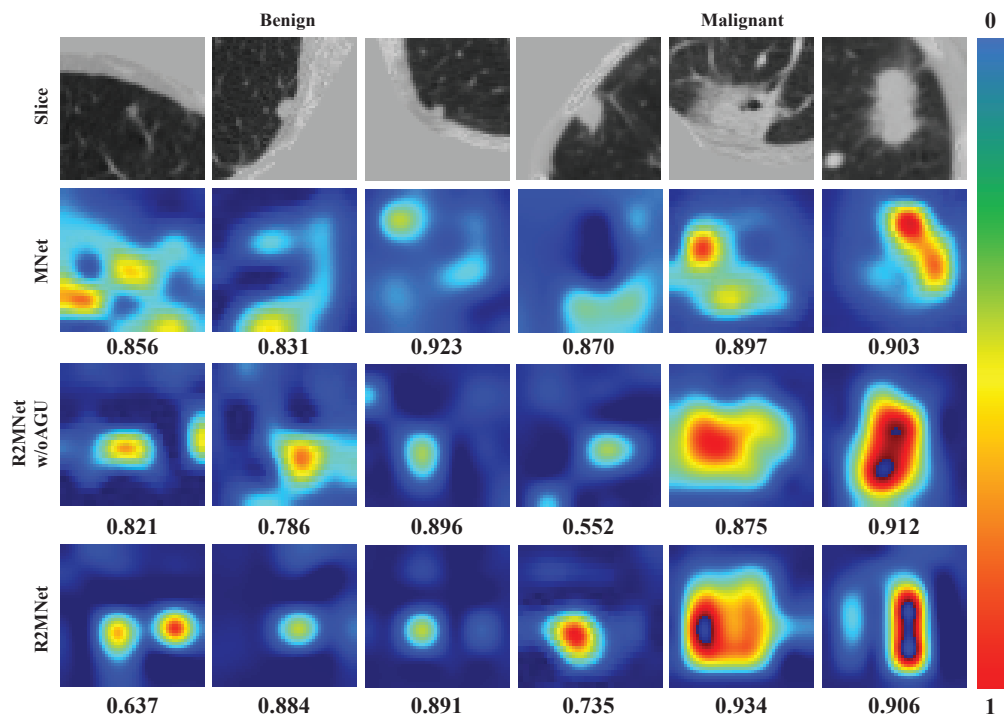}
\caption{Visualization of CDAM features derived from MNet, R2MNetw/oAGU, and R2MNet regarding malignant evaluation, respectively. The value under each sub-figure is probability predicted by the corresponding model. Note that we show the central slice only for visualizing convenience. Figure best viewed in color.}
\label{fig6}
\end{figure*}

\subsection{Model interpretation}
\begin{figure*}
\centering
\includegraphics[width=4.5 in]{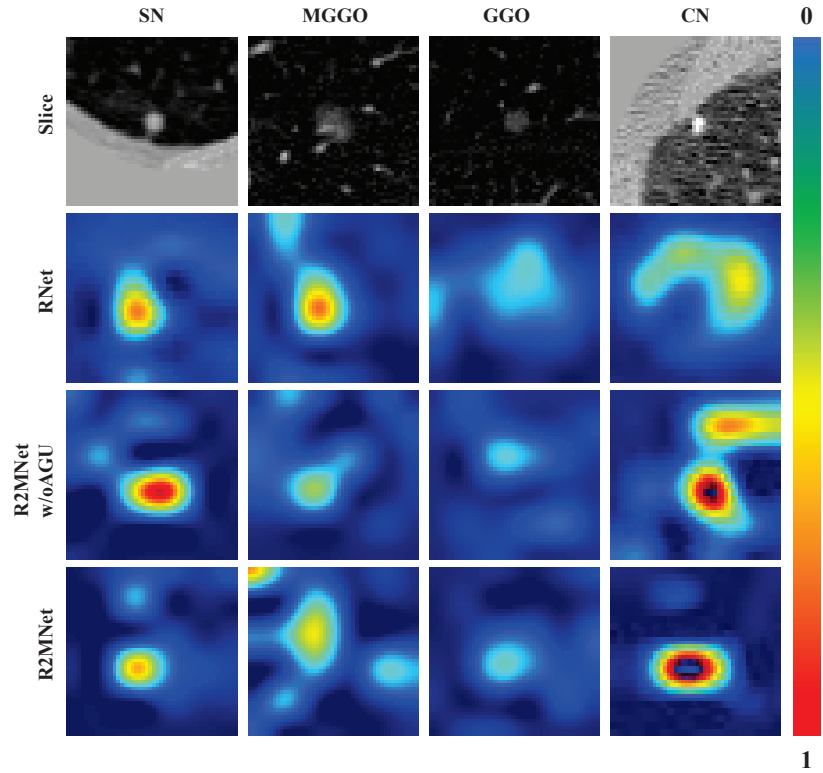}
\caption{Visualization of CDAM features derived from RNet, R2MNet\_w/oAGU, and R2MNet concerning radiology analysis, respectively. Note that we show the central slice only for visualizing convenience. Figure best viewed in color.}
\label{fig7}
\end{figure*}

Direct approaches that classify pulmonary nodule as benign or malignant from input CT data to the malignant probabilities lack of interpretation. To build explainable models, we provided visual explanations using the proposed CDAM. The experiments were performed both on malignant evaluation and radiology analysis to investigate voxel-wise importance regions which the models focus on in different tasks. Specifically, we employed the feature maps with a size of $256 \times 6 \times 6 \times 6$ after the last residual block in our model as activation maps. Since the activation maps are volume data, we adopted the center slice for visualization convenience. Fig.\ref{fig6} shows the CDAM features and its corresponding probabilities of MNet, R2MNetw/oAGU, and R2MNet concerning nodule malignant evaluation, respectively. The value below each sub-figure is the probability predicted by the corresponding model. Besides, we illustrated the CDAM features with respect to nodule radiology analysis in Fig.\ref{fig7}.

\section{Discussion}
\label{sec:discussion}
Automatic pulmonary nodule malignancy evaluation is an essential component of a CAD system for lung cancer diagnosis. Deep learning-based methods have demonstrated promising results on this task. Table \ref{table4} summarizes the related works from the literature. Shen et al. introduced a Multiscale CNN for nodule malignancy diagnosis and achieved an accuracy of $86.84\%$ on a selected LIDC-IDRI dataset \cite{MultiScaleCNN}. Nibali et al. adopted ResNet with multiview inputs for benign/malignant classification \cite{nodule_resnet}. They evaluated their method on the dataset derived from the LIDC-IDRI and achieved an accuracy of $89.90\%$. Al-Shabi et al. employed non-local blocks to model nodule global features and residual blocks to capture local features of nodule \cite{nodule_nonlocal}. They estimated the model on the selected LIDC-IDRI database with accuracy of $88.46\%$. However, classifying lung nodules as benign or malignant directly from the CT volume (or slice) lack clinical basis and explanations of the features extracted by the CNN. Therefore, the results are short of confidence level. Hussein et al. empirically established the significance of different high-level nodule attributes for malignancy determination \cite{tumornet}. Furthermore, they adopted CNNs to learn a series of features for nodule attributes then fused these features to predict the malignancy of pulmonary nodule in a multi-task learning manner \cite{hussein2017risk}. Similarly, Wu et al. proposed a multi-task learning CNN that integrated pulmonary nodule segmentation attributes and malignancy prediction \cite{wu2018joint}. Their approach simultaneously predicted the malignancy of lung nodules, segmented the nodule areas and learned nodule attributes, and aimed to tackle the problem of model interpretability.
Note that it can be difficult to pursue an objective cross-study comparison due to the differences in datasets, initialization methods, and experimental settings.

\begin{table}
\centering
\caption{Overview of previous methods for pulmonary nodule evaluation. Abbreviations: Information Processing in Medical Imaging (IPMI), International Symposium on Biomedical Imaging (ISBI), International Journal of Computer Assisted Radiology and Surgery (IJCARS).}
\begin{tabular}{ll}
\hline
Methods               & Accuracy \\ \hline
MCNN \cite{MultiScaleCNN}, IPMI                  & 86.84\%  \\
TurmorNet \cite{tumornet}, ISBI             & 82.47\%  \\
TurmorNet (Attributes) \cite{tumornet}, ISBI & 92.31\%  \\
Nodule-ResNet \cite{nodule_resnet}, IJCARS         & 89.90\%  \\
MIT-3DCNN \cite{hussein2017risk}, IPMI             & 91.26\%  \\
PN-SAMP \cite{wu2018joint}, ISBI               & 97.58\%  \\
Local-Global Networks \cite{nodule_nonlocal}, IJCARS & 88.46\%  \\
R2MNet, ours   & 94.74\%  \\ \hline
\end{tabular}
\label{table4}
\end{table}

Our method leveraged radiological features as a channel descriptor to assist lung nodule evaluation in a multi-task learning manner. Specifically, Table \ref{table2} indicates the results of the radiological analysis. Although radiology analysis is a auxiliary component in the R2MNet, R2MNet increased the accuracy among four nodule categories and the AUC score compared with RNet. Moreover, the ROC curves in Fig.\ref{fig4}(b) where the curve of R2MNet nearly surrounds that of RNet illustrate that the classification performance of R2MNet better than that of RNet. In nodule malignancy evaluation, Fig.\ref{fig4}(c) depicts the ROC curves of MNet and R2MNet in which the curves of the latter higher than that of the former. As indicated in Table \ref{table3}, in general, joint learning of radiology analysis and malignancy evaluation improved the performance compared to each individual. Combined learning facilitates communication between different tasks. We can conclude that these two tasks reinforce each other. Furthermore, comparing R2MNetw/oAGU with MNet, we can view the accuracy and AUC gain both in the two tasks. The performance of R2MNet on radiological analysis nearly equal to that of R2MNetw/oAGU. It is reasonable because the AGU module adopted radiological features to facilitate nodule malignancy evaluation. Indeed, the performance gain was obtained by R2MNet in malignancy estimation. On the other hand, Fig.\ref{fig5} depicts the box plots with average values and data distribution. The accuracy scores and AUC scores increase gradually among MNet, RNet,  R2MNetw/oAGU, and R2MNet, which further proves the effectiveness of the proposed methods. Viewing the boxes of MNet/RNet and R2MNetw/oAGU, one can conclude that although multi-task learning can bring performance gain, the results tend to fluctuate due to the hard convergence of the networks. However, the results of R2MNet are stable compared to others because introducing AGU into R2MNetw/oAGU enables the R2MNet to employ radiological features and then improve the adaptability of the model to different data.

Although performance improvement is one of a great purpose in developing deep learning-based methods, interpretability is essential as well. According to the experiences of radiologists, the shape and density of nodule regions are two critical factors that influence a nodule to be inferred as malignant. Fig.\ref{fig6} shows the CDAM features of MNet, R2MNetw/oAGU, and R2MNet concerning nodule malignant evaluation, respectively. MNet tended to be disturbed by the background noise and confused with benign and malignant features. In contrast, both R2MNet and R2MNetw/oAGU can focus on nodule regions except that they yielded a wrong identification in the first benign nodule. Furthermore, these two architectures paid higher attention to malignant nodules and lower attention to benign ones, which conforms to the risk of the nodules. According to the last two columns of benign and malignant nodules in Fig.\ref{fig6}, even though the MNet generated high probabilities, similar to other models, the concerning regions of MNet slightly deviate from the ground truth. On the contrary, R2MNet predicted low scores when it falsely located the nodule region, whereas MNet still generated high probability (Fig.\ref{fig6}, the first column). We can conclude that incorporating malignancy evaluation with radiology analysis can render the network emphasize nodule regions and characterize the shape and density features of nodules. 
Besides, the density of nodules plays a key role for nodule radiology analysis. As shown in Fig.\ref{fig7}, even though the four classes of nodules have different densities, the boundaries among them are confused, which led both the RNet and R2MNetw/oAGU to locate the nodule regions inaccurately. On the contrary, the R2MNet accurately located the nodules and lay different emphasis on these regions according to their densities, conforming with the clinical basis. Therefore, we can conclude that even though the results of R2MNet and R2MNetw/oAGU are similar, the inference process of R2MNet is more reasonable.

A major limitation of this work is that the input data depend on pulmonary nodule detection. The input data are derived from either manually choosing by radiologists or automatic detection by nodule detectors. Previous researches integrated multi-models into a synthetic system whose components were trained separately to performed different tasks. For example, Bonavita et al. developed a lung cancer classification pipeline that integrated a 3D CNN with an existing nodule detection framework \cite{bonavita2020integration}. Liao et al. adopted a 3D Faster R-CNN for patch-based nodule detection and integrated the leaky noisy-OR model into neural networks to solve lung cancer prediction \cite{dsb2017first}. Similarly, Zhu et al. build a DeepLung system to identify suspicious nodules and predict nodule malignancy \cite{deeplung}. Ozdemir et al. introduced a CAD system that included two sub-systems for nodule candidates segmentation and malignancy prediction \cite{3dprobabilisticCAD}. An end-to-end explainable CAD system for lung cancer diagnosis that integrates nodule detection,segmentation and malignancy prediction is of extensive clinical application value. This will be considered as our future work.

\section{Conclusion}
\label{sec:conclusion}
In this paper, we proposed the R2MNet that evaluated pulmonary nodule malignancy resorting to radiology analysis instead of directly infer malignant probability, which conformed to the clinical diagnosis procedure and increased the confidence of prediction results. Specifically, the radiological features were transformed into a channel descriptor that emphasized the informative malignant features and suppressed the less useful ones, so that the network could estimate the malignant risk based on radiological characteristics as did an experienced doctor to a patient. Besides, model explanations with CDAM shed light on the voxel-wise nodule regions which CNNs focussed on when they estimated nodule malignancy risk. The experimental results on the LIDC-IDRI database demonstrated the effectiveness of the proposed R2MNet.

\printcredits

\section*{Acknowledgement}
This work was supported by the Natural Science Foundation (Grant No. 2020J01472) and Provincial Science and Technology Leading Project (Grant No.2018Y0032) of Fujian Province, China. This work was also supported by Fujian Key Laboratory of Cardio-Thoracic Surgery (Fujian Medical University)

\bibliographystyle{model1-num-names}
\bibliography{r2mnet_bib}

\end{document}